\documentstyle[prl,twocolumn,aps,epsf]{revtex}
\begin{document}
\twocolumn[\hsize\textwidth\columnwidth\hsize\csname @twocolumnfalse\endcsname
\draft
\title{Contrarian Deterministic Effect: the ``Hung Elections Scenario"}
\author{Serge Galam}
\address{Laboratoire des Milieux D\'esordonn\'es et H\'et\'erog\`enes,
CNRS UMR 7603, Universit\'{e} Paris 6,\\
4, place Jussieu, 75252 Paris Cedex 05, France}
\date{galam@ccr.jussieu.fr}
\maketitle
\begin{abstract}
A contrarian is someone who deliberately decides to oppoe the prevailing
choice of others. The Galam model of two state opinion dynamics 
incorporates agent updates
by a single step random grouping in which all participants adopt the opinion of
their respective local majority group. The process is repeated until
a stable collective state
is reached; the associated dynamics is fast. Here we show that the 
introduction of
contrarians may give rise to interesting dynamics generated phases 
and even to a critical
behavior at a contrarian concentration $a_c$.
For $a<a_c$ an ordered phase
is generated with a clear cut majority-minority splitting. By 
contrast when $a>a_c$
the resulting disordered phase has no majority: agents keep shifting 
opinions but no
symmetry breaking (i.e., the appearance of a majority) takes place.
Our results are employed to explain the outcome of
the 2000 American presidential elections and that of the 2002 German 
parliamentary elections.
Those events are found to be inevitable. On this basis the ``hung 
elections scenario''
is predicted to become a common occurrence in modern democracies.
\end{abstract}
]


In this letter, we study the effects of contrarian choices on the dynamics of
opinion forming. A contrarian is someone who deliberately decides to oppose
the prevailing choice of others whatever this choice is 
\cite{contra}. Contrarian
strategy is becoming a growing new trend of modern democracies most studied
in finance \cite{contra,corcos}.

The Galam model of two state opinion
dynamics incorporates
agent updates by a series of single steps. In each step random groups 
are formed in which
all participants adopt the opinion of their respective local majority group
\cite{mino,chopard1,voting-all}.
The process is repeated until a stable collective state
is reached. The associated dynamics is fast and leads to a total 
polarization along either
one of the two competing states A and B. The direction of the opinion 
flow is monitored by
an unstable separator at some critical density $p_c$ of agent
supporting the A opinion.

In the case of odd size groups, $p_{c}=\frac{1}{2}$. By contrast even sizes
make $p_{c}\neq \frac{1}{2}$. The corresponding
asymmetry in the dynamics of respectively opinion A and B arises from 
the existing
of local collective doubts at a tie. The unstable separator may then 
be simultaneously
at a value of $23\%$ for one state and at $77\%$ for the other \cite{chopard1}.
Large size group accelerates reaching
the final state with a drastic reduction in the number of required 
updates. In the limit
of a single grouping which includes the entire population one update 
is enough to complete the full
polarization.
Recently a generalization to any distribution of group sizes was achieved
yielding a very rich and complex phase diagram \cite{mino}.
The model was subsequently applied to rumor phenomena \cite{rumor}.

Earlier version of this approach is found in the study of voting in democratic
hierarchical systems \cite{voting-all}. There, groups of agents vote for a
representative to the higher
level using a local majority rule. In the mean field limit, going up 
the hierarchy turns
out to be exactly
identical to an opinion forming process in terms of equations and dynamics.
Instead of voting, agents update
their opinions. The probability of electing an A representative at 
some hierarchy level
$n$ is equal to the proportion of A opinions after $n$ updates 
\cite{voting-all,chopard1}.
Recent studies by Krapivsky and Redner further explored the dynamical 
properties
of the Galam model, restricted to one group of size 3 \cite{redner1}.

This work contributes to the now growing field of applications of
Statistical Physics to social and political behaviors
\cite{strike,vicsek,sorin,deffuant,vespignani,frank,weron,stauffer}.
First denoted ``Sociophysics" in a founding paper \cite{strike}
we extend the label to ``Global Physics". At this stage
it is worth stressing we are not aiming at an exact description of 
the real social
and political life, but rather, doing some crude approximations, to enlighten
essential features of an otherwise very complex and multiple phenomena.

Here, the dynamics of Contrarian behavior is studied using the Galam 
model of two state
opinion dynamics restricted to odd sizes. Introduction of contrarians 
at a low density $a$
is found to unfold the total polarization dynamics. The corresponding 
fully ordered
state with one unique opinion becomes mixed with a stable 
majority-minority splitting.
But the symmetry breaking is preserved with a clear cut majority 
along the initial global
majority. The unstable separator is also left unchanged at $p_{c}=\frac{1}{2}$.

However, contrarians are found to give rise to a critical
behavior at a contrarian concentration $a_c$. When $a>a_c$
a new disordered stable phase with no majority appears. There agents 
keep shifting opinions
but no symmetry breaking (i.e., the appearance of a majority) takes place.
Contrarians have turned the unstable separator $p_c$ into the
unique stable attractor of the dynamics. Opinion flows ahve been reversed.
The value of $a_c$ depends on the size distribution of update groups.

Our results are employed to explain the outcome of
the 2000 American presidential elections and that of the 2002 German 
parliamentary elections.
Those events are found to be inevitable. On this basis the ``hung 
elections scenario"
is predicted to become a common occurrence in modern democracies.

We start with a very simple model of opinion forming \cite{voting-all,mino}.
Considering an ideal society before a major election, people start 
discussing the issue
during the election campaign. Groups are formed randomly in which all 
participants adopt
the local majoirity state. Focusing first on the group size 3,
an initial 2 A (B) with one B (A) ends up with 3 A (B).
To follow
the time evolution of the vote intentions we need an estimate of
the numbers of respective vote intentions $N_+(t)$ for A and $N_+(t)$
for B  at some time t from a N person population. It can be evaluated 
using polls.
Each person is assumed to have
an opinion with $N_+(t)+N_-(t)=N$. Corresponding individual probabilities to a
vote intention in favor of A or B writes,
\begin{equation}
p_{\pm}(t)\equiv \frac{N_{\pm}(t)}{N}  ,
\end{equation}
with,
\begin{equation}
p_+(t)+p_-(t)=1 .
\end{equation}
Accordingly, one cycle of local opinion updates via three persons 
grouping leads
to a new distribution of vote intention as,
\begin{equation}
p_+(t+1)=p_+(t)^3+3p_+(t)^2p_-(t) ,
\end{equation}
where $p_+(t+1)>p_+(t)$  if $p_+(t+1)>\frac{1}{2}$ and $p_+(t+1)<p_+(t)$
if $p_+(t+1)<\frac{1}{2}$. Indeed from Eq. (2) vote intention $p_+(t)$  flows
monotonically
toward either one of two stable point attractors at $P_{+A}=1$ and $P_{+B}=0$.
An unstable point
separator attractor is located at $p_c=\frac{1}{2}$. It separates the 
two basins
of attraction
associated respectively to the point attractors.

During an election campaign people go trough several successive different local
discussions. To follow the associated vote intention evolution we 
iterate Eq. (2). A number
of m discussion cycles gives the series $p_+(t+1), p_+(t+2)... 
p_+(t+m)$. For instance
starting at $p_+(t)=0.45$  leads successively after 5 intention 
updates to the series
$p_+(t+1)=0.43, p_+(t+2)=0.39, p_+(t+3)=0.34, p_+(t+4)=0.26, p_+(t+5)=0.17$
with a continuous
decline in A vote intentions. Adding 3 more cycles would result in zero A vote
intention with $p_+(t+6)=0.08, p_+(t+7)=0.02 and p_+(t+8)=0.00$. 
Given any initial
intention vote distribution, the random local
opinion update leads toward a total polarization of the collective opinion.
Individual and collective opinions stabilize simultaneously along the same and
unique vote intention either A or B.

The update cycle number to reach either one of the two stable attractors can be
evaluated from Eq. (2). It depends on the distance of the initial 
densities from
the unstable point attractor. An analytic formula is derived below (see Eq. 6).
However, every update cycle takes some time length, which may 
correspond in real terms
to some number of days.
Therefore, in practical terms the required time to eventually complete the
polarization process is much larger than the campaign duration, thus preventing
it to occur. Accordingly, associate elections never take place at the stable
attractors. From above example at $p_+(t)=0.45$ , two cycles yield a result of
$39\%$ in favor of A and $61\%$ in favor of B. One additional update 
cycle makes
$34\%$ in favor of A and $66\%$ in favor of B.

At this stage we are in a position to insert in the model the existence of
contrarians. A contrarian is defined as
follows \cite{contra}. Once a local group reaches a consensus driven 
by the majority rule,
there exists some people, which once they left the group, shift to the opposite
vote intention. The shift is independent of the choice itself. 
Setting contrarian
choices at a density $a$ with $0\leq a\leq 1$ , the density of A opinion
given by Eq. (2)
becomes,
\begin{eqnarray}
p_+(t+1) & = & (1-a)[p_+(t)^3+3p_+(t)^2p_-(t)]\\
\nonumber& &
+a[p_-(t)^3+3p_-(t)^2p_+(t)] ,
\end{eqnarray}
where first term corresponds to the regular update process and second 
term to contrarian
contribution from local groups where the local majority was in favor of B.
 From Eq. (4), the effect of low-density contrarian choices is readily seen as
illustrated in Figure (1) in the case $a=0.10$, i.e., with $10\%$ 
contrarian choices
as compared to the pure case $a=0$.

\begin{figure}
\epsfxsize=\columnwidth
\begin{center}
\caption{Equation (4) with $P_{+}(t+1)$  as function of $P_{+}(t)$ at 
respectively
$a=0$ and $a=0.10$.
In the second case the two stable point attractors have moved from total
polarization towards coexistence of mixed vote intentions with a clear cut
majority-minority splitting.
}
\vspace{2mm}
\centerline{\epsfbox{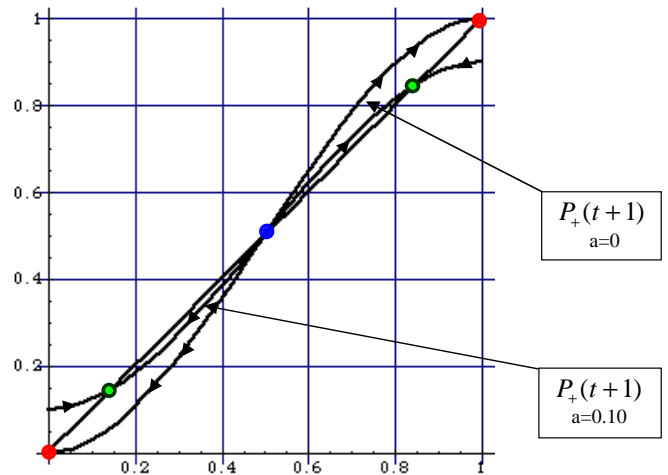}}
\end{center}
\end{figure}

Main effects are twofold. First both stable point attractors are shift toward
coexistence vote intention values. Total polarization is averted with,
\begin{equation}
P_{+A(B)}=\frac{(2a-1)\pm\sqrt{12a^2-8a+1}}{2(2a-1)},
\end{equation}
which are defined only in the range  $a\leq\frac{1}{6}$. For instance a value
of $a=0.10$ yields $P_{+A}=0.85$ and $P_{+B}=0.15$. At $P_{+A}=0.85$  exists
a stable coexistence of vote intentions at respectively $0.85$ in A favor
with $0.15$ in B favor. The reverse holds at $P_{+B}=0.15$. At 
contrast contrarian
choices keep unchanged the unstable point separator at $\frac{1}{2}$.

The second effect from contrarian choices is an increase in the number of
cycle updates in reaching the stable attractors. For instance starting as above
at $p_+(t)=0.45$  with $a=0.10$ leads now to the series $p_+(t+1)=0.44,
p_+(t+2)=0.43, p_+(t+3)=0.42, p_+(t+4)=0.40, p_+(t+5)=0.38$. Additional $12$
updates are required to reach the
stable attractor at $0.15$. All cycles score to $17$ against only $8$ without
contrarian
choices. A vote at two update cycles from above same example would 
give a voting
result of $43\%$ in favor of A and $57\%$ in favor of B instead of 
respectively $39\%$
and $61\%$ at $a=0$.

An approximate formula can be derived from Eq. (4) to evaluate the 
update cycle number
required to reach either one of the two stable attractors. It writes,

\begin{equation}
n\simeq \frac{1}{\ln[\frac{3}{2}(2a-1)]}\ln[\frac{p_c-P_S}{p_c-p_+(t)}]
+\frac{1.85}{(2a-1)^{5.2}} ,
\end{equation}
where last term is a fitting correction. $P_S=P_{+B}$ if $p_+(t)<p_c$
while $P_S=P_{+A}$ when $p_+(t)>p_c$. The number of
cycles being an integer, its value is obtained from Eq. (6) rounding to
an integer. At $a=0$, i.e., no contrarian choices, $n$ is always a small number
as shown in Figure (2). Eq. (6) gives $8$ at an initial value $p_+(t)=0.45$
and $4$ at $p_+(t)=0.30$,
which are the exact values obtained by successive iterations from Eq. (3).
At $a=0.10$ we found also the exact values of $17$ and $9$ as from Eq. (4).

\begin{figure}
\epsfxsize=\columnwidth
\begin{center}
\caption{Approximate number of cycles of vote intention updates to 
reach a total
polarization of opinion as function of an initial support  $P_{+}(t)$.
}
\vspace{2mm}
\centerline{\epsfbox{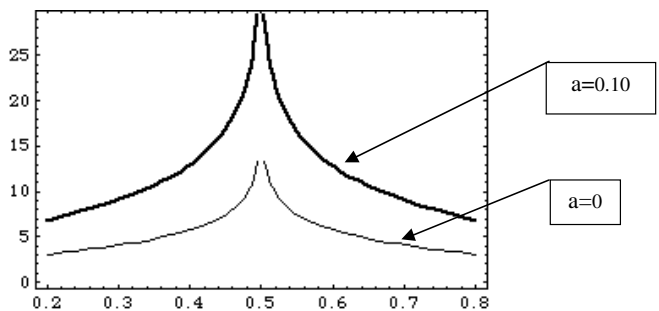}}
\end{center}
\end{figure}

Both Eq. (6) and Figure (2) show explicitly the contrarian choice drastic
effect in increasing the number of required levels to reach the stable point
attractors. That means much longer real time. In practical terms it implies
a quasi-stable coexistence of both vote intentions not too far from fifty
percent but yet with a clear-cut majority in one direction, which is determined
by the initial majority.

However contrarian choices may lead to a radical qualitative change in the
whole vote intention dynamics. Eq. (5) shows that at a density of
$a=\frac{1}{6}\simeq 0.17$ , contrarian
choices make both point attractors to merge simultaneously at the
unstable point separator  $p_c=\frac{1}{2}$ turning it to a stable 
point attractor.
Consequences on the vote intention dynamics are drastic. The flow
direction is reversed making any initial densities to converge
toward a perfect equality between vote intention for A and B. In 
physical terms,
contrarians produce a phase transition from a majority-minority phase 
into a fifty
percent balance phase with no majority-minority splitting.
In the ordered phase elections
always yield a clear-cut majority. At contrast in the disordered 
phase elections lead
to a random outcome driven by statistical fluctuations.
An illustration is shown in Figure (3) for $20\%$ of contrarians.

\begin{figure}
\epsfxsize=\columnwidth
\begin{center}
\caption{$P_{+}(t+1)$ as function of $P_{+}(t)$ at $a=0$ and 
$a=0.20$. In the first
case the vote intention
flows away from the unstable point attractor at $\frac{1}{2}$ toward 
either one of the stable
point attractors at zero or one. In the second case, contrarian choices have
reversed the flow directions  making any initial densities to flow
toward $\frac{1}{2}$, the now stable and unique point attractor.
}
\vspace{2mm}
\centerline{\epsfbox{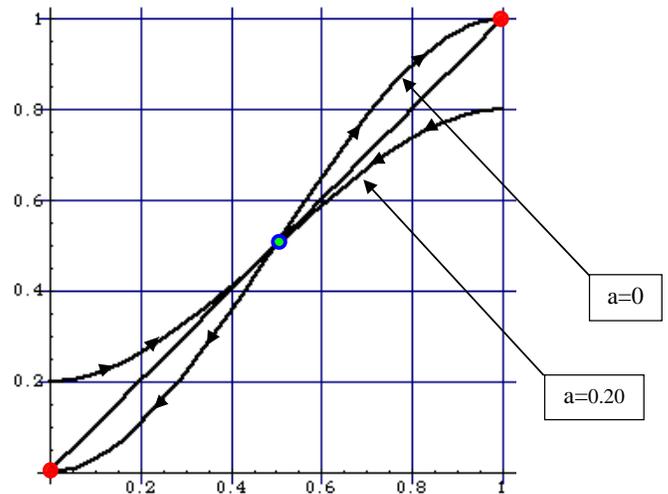}}
\end{center}
\end{figure}

In real social life people don't meet only by group of 3. However, generalizing
above approach to larger sizes is straightforward and does not change the
qualitative feature of the model. Dynamics reversal driven by contrarians
towards the disorder phase with no majority-minority splitting is preserved.
The main effect is an
increase in the value of the contrarian critical density at which the phase
transition occurs. In the case
of an odd size k, Eq. (4) becomes,
\begin{equation}
p_+(t+1)=(2a-1)
\sum_{i=\frac{k+1}{2}}^k  C_k^i p_+(t)^i p_-(t)^{(k-i)}+a ,
\end{equation}
where $C_k^i\equiv \frac{k!}{(k-i)!i!}$ . The instrumental parameter 
in determining
the flow direction and the
associate phase transition is the eigenvalue at the point attractor 
$p_c=\frac{1}{2}$.
It is given by,
\begin{equation}
\lambda=(2a-1)\left[\frac{1}{2}\right]^{k-1}\sum_{i=\frac{k+1}{2}}^k 
(2i-k)C_k^i .
\end{equation}

The range  $\lambda >1$ determines an unstable point attractor with 
an ordered phase
characterized by the existence of a majority-minority splitting.
At contrast, $\lambda <1$ makes the
point attractor stable. The case  $\lambda =1$ determines the 
critical value of the
contrarian choice density $a_c$  at which the phase transition occurs.
 From Eq. (8), we get,
\begin{equation}
a_c=\frac{1}{2} \left( 1-\left[ (\frac{1}{2})^{k-1}
\sum_{i=\frac{k+1}{2}}^k (2i-k)C_k^i\right]^{-1} \right) .
\end{equation}
In the case $k=3$  we recover the above result $a=\frac{1}{6}\simeq 0.17$.
 From Eq. (9) it is seen that $a_c\rightarrow \frac{1}{2}, k 
\rightarrow +\infty$
with $0.33$ at $k=5$ and $0.30$ at $k=9$.

We have presented a simple model to study the effect of contrarian choices
on opinion forming. At low densities $a$ the opinion dynamics leads 
to a mixed phase with a
clear cut majority-minority splitting. However, beyond some critical
density $a_c$, contrarians make all the attractors to merge at the 
separator $p_c$. It
becomes the unique attractor of the opinion dynamics. When $a>a_c$ 
vote intentions flow
deterministically with time towards an exact equality between A and B 
opinions. In this new
disordered stable phase no majority appears. Agents keep shifting opinions
but no symmetry breaking (i.e., the appearance of a majority) takes place.
There an election would result in effect in a random winner due to 
statistical fluctuations.
The value of $a_c$ depends on the size distribution of update groups.

Accordingly, our results shed a totally new light on recent elections in
America (2000) and Germany (2002). It suggests those ``hung elections" were not
chance driven. On the opposite, they are a deterministic outcome
of contrarians. As a consequence, since contrarian
thinking is becoming a growing trend of modern societies, the subsequent
``hanging chad elections'' syndrome is predicted to become both inevitable
and of a common occurrence.

While finalizing this manuscript we have notice Ref. \cite{redner2} by Mobilia
and Redner in which
a phase transition in a disordered opinion phase is also obtained via
an interesting extension of Galam model (restricted to one group of size 3)
which combines locally majority and minority rules. However the 
microscopic rules used as
well as the socio-political interpretation and the the critical 
values are different
from those of the present work.

\section*{Acknowledgment}

I would like to thank Yuval Gefen for helfull comments on the manuscript.


\end{document}